\documentclass[prl,twocolumn,showpacs,amsmath,amssymb]{revtex4}

\usepackage{graphicx,dcolumn,bm}

\begin{document}

\title{Anisotropic states of two-dimensional electrons in high magnetic fields}

\author{A.M. Ettouhami} 
\email{mouneim@physics.utoronto.ca}
\affiliation{Department of Physics, University of Toronto, 60 St. George St., Toronto M5S 1A7, Ontario, Canada}

\author{C.B. Doiron}
\affiliation{Department of Physics and Astronomy, University of Basel,
Klingelbergstrasse 82, CH-4056 Basel, Switzerland}

\author{F.D. Klironomos}
\affiliation{Department of Physics, University of California, Riverside, CA 92521}

\author{R. C\^ot\'e} 
\affiliation{D\'epartement de Physique, Universit\'e de Sherbrooke, Qu\'ebec, Canada, J1K 2R1}

\author{Alan T. Dorsey}
\affiliation{Department of Physics, University of Florida, P.O. Box 118440, Gainesville, FL 32611}

\date{\today}

\begin{abstract}

We study the collective states formed by two-dimensional electrons in Landau levels of index 
$n\ge 2$ near half-filling. By numerically solving the self-consistent Hartree-Fock (HF)
equations for a set of oblique two-dimensional lattices, we find that the stripe state 
is an anisotropic Wigner crystal (AWC), and determine its precise structure
for varying values of the filling factor.
Calculating the elastic energy, we find that the shear modulus of the AWC is small but finite 
(nonzero) within the HF approximation. This implies, in particular, that the 
long-wavelength magnetophonon mode in the stripe state vanishes like $q^{3/2}$
as in an ordinary Wigner crystal, and not like $q^{5/2}$ as was found in previous studies
where the energy of shear deformations was neglected.

\end{abstract}



\maketitle

\begin{figure}[t]
\includegraphics[width=8.09cm, height=5cm]{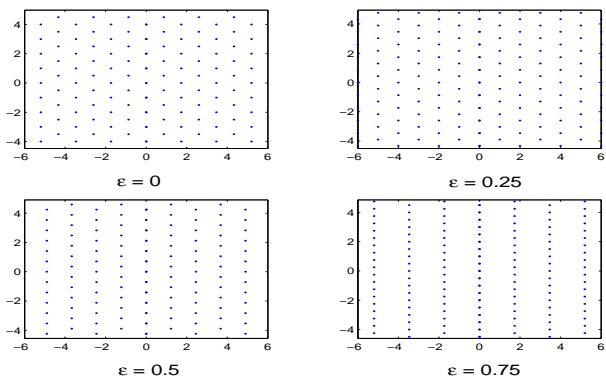}
\caption[]{(Color online) 
Evolution of the WC appearance for various values of the anisotropy parameter $\varepsilon$ of
Eq. (\ref{Rnm}). For $\varepsilon\geq 0.5$, the lattice shows a pronounced channel-like structure, 
with each channel consisting of a one-dimensional periodic chain of electron guiding centers.
}\label{Fig:AWC}
\end{figure}

One of the most fascinating properties of two-dimensional electrons in a strong perpendicular
magnetic field is the strong anisotropy these systems display around half filling
of Landau levels (LL's) of index $n\geq 2$. This anisotropy, first seen in DC 
transport measurements \cite{Lilly1999,Du1999}, has been interpreted 
early on \cite{Koulakov1996,Moessner1996} as evidence for the existence of a unidirectional charge density wave,
termed ``stripe state", near half filling, whereas the insulating states observed for small current
at intermediate and small partial filling factor $\nu^*=\nu-2n$ ($\nu$ being the total filling factor)
were identified as pinned crystalline states of the 2D electron system 
\cite{Koulakov1996,Moessner1996,Cote2003,Goerbig2004,Haldane2000,Shibata2001}.
While the concept of a stripe phase, with the density of electron guiding centers taking alternating 
{\em uniform} values that are piecewise constant along a periodic stripe structure, is sufficient to
explain the anisotropy of the DC response, other structures may lead to an anisotropic response as well.
One such structure is an {\em anisotropic} Wigner crystal (AWC), whose existence was suggested 
by microscopic \cite{MacDonald2000} as well
as Hartree-Fock (HF) \cite{Cote2000,Cote2003} studies, and which under certain circumstances (see below)
resembles a stripe structure, but with a periodic modulation of the electronic density along the stripes. 
In this Letter, we report on a systematic study of the cohesive energy of anisotropic triangular WCs, 
with special focus on the LL $n=2$. By numerically solving the self-consistent HF equations, we find that the 
stripe state is in fact an AWC. Indeed, as it can be seen in Fig. \ref{Fig:AWC},
a highly anisotropic rhombic lattice may be viewed as
a periodic arrangement of one-dimensional stacks of equidistant electron guiding centers (see Fig. \ref{Fig:AWC}).
Calculating the elastic moduli of the AWC, we find that the energy of shear deformations
corresponding to channels sliding past each other, although very small, does {\em not} vanish.
This has the important consequence that the stripe state behaves as a smectic on short length scales,
but on scales longer than a crossover length $L_{cr}$ (that we estimate below, see Eq.(\ref{Eq:Lcr}))
it behaves like an ordinary two dimensional anisotropic solid.


\begin{figure}[t]
\includegraphics[width=8.09cm, height=7cm]{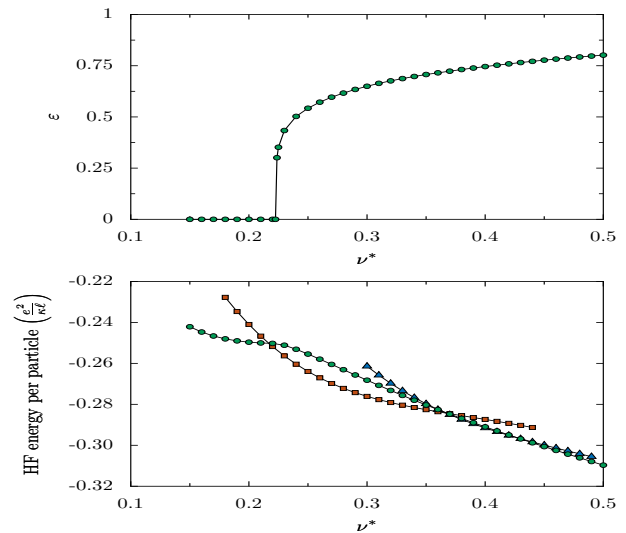}
\caption[]{(Color online) 
Upper panel: values of the anisotropy parameter $\varepsilon$ that minimize 
the cohesive energy of the WC vs. $\nu^*$.
Lower panel: cohesive energies of the Wigner crystal (circles), 2e bubble solid (squares),
and 3e bubble solid (triangles). For the WC, $\varepsilon$ values from the upper panel
are used. Both panels are for LL $n=2$. 
}\label{Fig-anisotropicHF}
\end{figure}

In order to probe anisotropic crystalline states and the corresponding 
cohesive energies of the 2D electron system, we have used the self-consistent Hartree-Fock
method of Ref. \cite{Cote2003}. In this work, we have systematically studied a family of oblique 
(rhombic) crystals parametrized by the following lattice vectors ($n$ and $m$ are integers):
\begin{eqnarray}
{\bf R}_{nm} = na_1 \,\hat{\bf x} + (2m+n)a_2\,\hat{\bf y} \,,
\label{Rnm}
\end{eqnarray}
where $a_1 = \sqrt{3}\,a/2\sqrt{1-\varepsilon}$ and $a_2=\sqrt{1-\varepsilon} a/2$, and
where $\varepsilon$ is a positive parameter (such that $0\leq \varepsilon <1$)
which quantifies the degree of anisotropy of the lattice (with $\varepsilon=0$ corresponding to
an isotropic triangular lattice), and $a=\ell \big(4\pi M/\sqrt{3}\,\nu^*\big)^{1/2}$ 
is the average lattice spacing of an isotropic lattice with $\varepsilon=0$ at the same density
($M$ is the number of electrons per bubble in a general bubble crystal).
At any fixed partial filling factor $\nu^*$, we compute the cohesive energy $E_{coh}(\varepsilon)$ 
for a sampling of values of $\varepsilon$ in the interval $0\leq \varepsilon < 1$, 
and determine the value of $\varepsilon$ that minimizes this last quantity. 
We find that the cohesive energy of bubble crystals with $M=2$ and $3$
is minimized by the value $\varepsilon=0$ throughout the whole $\nu^*$ range.
For the Wigner crystal (with $M=1$) however, we find that the cohesive energy
is minimized by nonzero values of $\varepsilon$ for $\nu^*\geq 0.22$.
The resulting variation $\varepsilon(\nu^*)$ for the WC is shown in the upper panel of 
Fig. \ref{Fig-anisotropicHF}, while the lower panel shows the resulting cohesive energies of the 
various phases. As it can be seen, the ground state of the 2D electron system goes through
a succession of {\em isotropic} ground states (Wigner crystal and bubble solids with $M=2$ and $3$) 
at small and intermediate values of the partial filling factor $\nu^*$.
At $\nu^*\simeq 0.42$, the system undergoes a first order phase transition
to a modulated stripe state, which in this case is nothing but an AWC, with values 
of $\varepsilon$ ranging from $\varepsilon\simeq 0.76$ at the transition to 
about $\varepsilon\simeq 0.80$ at half filling in LL $n=2$.

Having investigated the phase diagram of the 2D electronic system, we now want to
investigate the elastic properties of the resulting crystalline phases. 
For the anisotropic triangular lattice we are considering here, 
the general theory of linear elasticity in two dimensions
dictates that the elastic energy density be of the form \cite{Ashcroft-Mermin}: 
\begin{align}
{\cal E}_{el} = \frac{1}{2}\,\big(
c_{11}u_{11}^2 + c_{22}u_{22}^2 + 4 c_{66}u_{12}^2 
+ 2c_{12}u_{11}u_{22} 
\big),
\label{Eq:E_el}
\end{align}
where $u_{\alpha\beta}=\frac{1}{2}(\partial_\alpha u_\beta + \partial_\beta u_\alpha)$
is the symmetric strain tensor, and where we have used reflection symmetry (in both $x$ and $y$) 
to reduce the number of elastic moduli from six to four. (For an isotropic triangular crystal, 
the number of elastic moduli can further be reduced by symmetry \cite{Ashcroft-Mermin},
so that the elastic energy can be written in terms of only two elastic constants, the compression
($c_{11}$) and shear ($c_{66}$) moduli.) Note that stability of two dimensional solids requires
that the shear modulus be strictly positive, $c_{66}>0$.
For two-dimensional uniform deformations, 
such that the displacement vector ${\bf u}({\bf r})$ is given by $u_\alpha = u_{\alpha,\beta} x_\beta$,
with constant coefficients $u_{\alpha,\beta}$, it can be shown \cite{Miranovic2001} that the reciprocal lattice also experiences 
a homogeneous deformation, with the deformed reciprocal lattice vectors $\{{\tilde{\bf Q}}\}$
given in terms of the original ones $\{{\bf Q}\}$ by
$\tilde{Q}_\alpha = Q_\alpha - u_{\beta,\alpha} Q_\beta + {\cal O}(u^2)$.
To extract a given elastic constant in Eq. (\ref{Eq:E_el}), we calculate the HF energy of the 
corresponding distorted state (with distortion amplitude $u_0$), 
and extract the elastic constant from the excess (elastic) energy ${\Delta}E_{el}(u_0)=E_{HF}(u_0)-E_{HF}(0)$.
For example, to extract the shear modulus $c_{66}$ 
we calculate the HF energy $E_{HF}(u_0)$ of the distorted crystal 
(which for small $u_0$ has an optimal anisotropy $\varepsilon$ that is very close to that of
the undistorted lattice) using the reciprocal lattice vectors
$(\tilde{Q}_x =  Q_x - u_0 Q_y\,,\tilde{Q}_y  =  Q_y)$, corresponding to the shear deformation 
polarized along $y$ such that ${\bf u}({\bf r}) = u_0 \, x \,\hat{\bf y}$. 
The shear modulus (in units of {\em energy/particle}) is then given by
$c_{66} = 2{\Delta}E_{el}(u_0)/u_0^2$ 
(to obtain the shear moduli in the conventional units of energy per unit area,
one should divide by the area of a unit lattice cell $A_c=2\pi M\ell^2/\nu^*$).
In our case where we consider electrons interacting through the long range Coulomb potential, the above procedure is appropriate 
only for the shear moduli (which can be shown to be only weakly dispersive \cite{Cote2005,Ettouhami2005}), 
and for the non-dispersive part of the compression moduli (the compression moduli
are well-known to be dominated by a $1/q$ plasmon contribution in the long wavelength limit).

\begin{figure}[t]
\includegraphics[width=8.09cm, height=5cm]{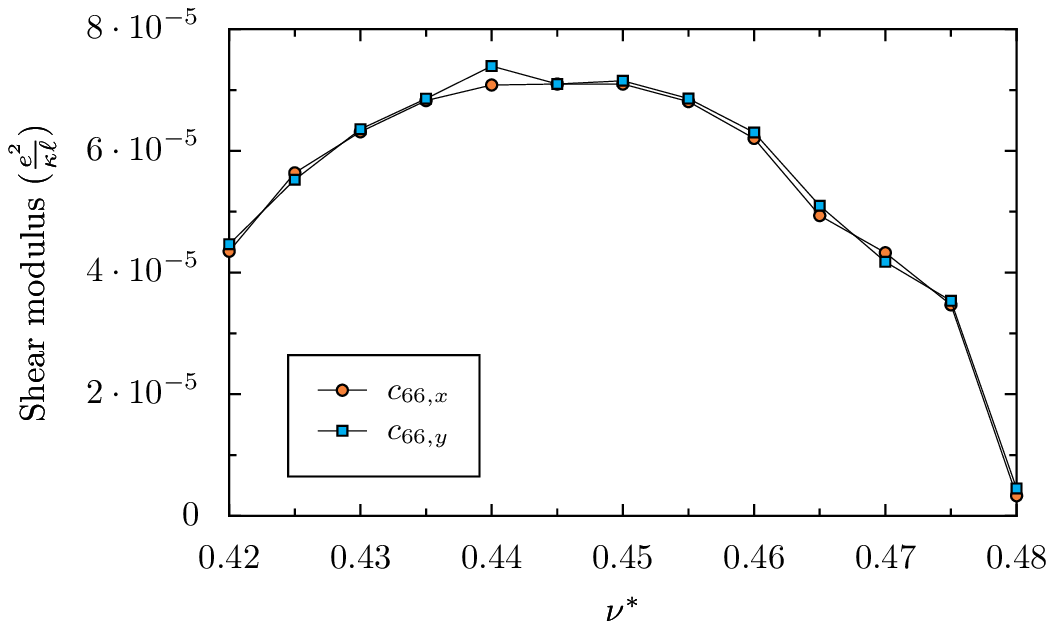}
\caption[]{(Color online) 
Shear modulus of the anisotropic Wigner crystal in $n=2$
in units of $e^2/\kappa\ell$.
The values of $c_{66}$ obtained for a shear polarized along $\hat{\bf x}$
and along $\hat{\bf y}$ are denoted by $c_{66,x}$ and $c_{66,y}$, respectively. 
}\label{Fig-anisotropic-moduli}
\end{figure}

We now turn our attention to the calculation of the shear modulus of the actual
anisotropic ground states of our system. We choose to do the calculation in two different ways,
by calculating the energy of shear deformations both along $x$ (${\bf u}({\bf r})=u_0y\hat{\bf x}$)
and along $y$ (${\bf u}({\bf r})=u_0x\hat{\bf y}$). In Fig. \ref{Fig-anisotropic-moduli} we plot the behavior 
of the two corresponding moduli, that we call $c_{66,x}$ and $c_{66,y}$ vs. $\nu^*$ for the AWC ground states we found
above in the $n=2$ LL. We first note that $c_{66,x}$ and $c_{66,y}$ are equal, as they should be, 
which confirms the fact that the shear deformations in our lattice are controlled by a single shear modulus $c_{66}$.
Comparing the values of this shear modulus $c_{66}$ obtained over the stripe phase to the values
obtained in Refs. \onlinecite{Cote2005,Ettouhami2005} for the Wigner crystal and bubble phases, 
we see that $c_{66}$ jumps discontinuously by over two orders
of magnitude from values of order $0.01(e^2/\kappa\ell)$ for $\nu^*\leq 0.42$ where the electronic solid is isotropic,
to values of order $5\times 10^{-5}(e^2/\kappa\ell)$ in the anisotropic regime $0.42\leq \nu^*\leq 0.5$
($\kappa$ is the dielectric constant of the host semiconductor).
This indicates that sliding the channels in Fig. \ref{Fig:AWC} past each other costs a finite, albeit very small shear 
energy. One may of course worry that the smallness of the shear moduli we find above is a numerical artifact
with no actual physical meaning. In the following, we shall argue that non-zero shear moduli are a natural
consequence of the modulated density along the stripes in the AWC.

Central to our discussion is the observation that there is a fundamental difference between our
AWC state, where electron guiding centers form {\em modulated} channels breaking
translational invariance in both $x$ and $y$ directions,
and the conventional stripe state, with a {\em uniform} density of guiding centers along the
stripes. For, while in the latter case, assuming the stripes are parallel to the $\hat{\bf y}$
direction, a deformation of the form ${\bf u}=u_0y\hat{\bf x}$ corresponds to a rigid {\em rotation} of
the whole stripe structure (with no overall energy cost), in our case with the electron guiding centers
forming modulated channels, the deformation ${\bf u}=u_0y\hat{\bf x}$ does {\em not} correspond to a pure
rotation of the system, and in fact involves a {\em compression} of the electron guiding centers along the
direction of the channels, which costs a nonzero elastic energy $\sim c_{66}(\partial_yu_x)^2$.
This point can be seen more clearly if we consider the following displacement field of the AWC,
${\bf u}_{nm} = -\theta(n+2m)a_2\hat{\bf x}$
which, in continuous notation, corresponds to the shear deformation ${\bf u}({\bf r})=-\theta y\hat{\bf x}.$
Performing an infinitesimal rotation of angle $\theta$ to the new cartesian frame $(x'y')$ such that
$\hat{\bf x}  =  \hat{\bf x}' - \theta\hat{\bf y}'$ and
$\hat{\bf y}  =  \theta\hat{\bf x}' + \hat{\bf y}'$,
we obtain that the displaced positions ${\bf r}_{nm}={\bf R}_{nm}+{\bf u}_{nm}$ are given in the $(x'y')$
frame by ${\bf r}_{nm} = na_1\hat{\bf x}' + [(n+2m)a_2 - na_1\theta]\,\hat{\bf y}'$.
As it can be seen, the shear deformation ${\bf u}_{nm} = -\theta(n+2m)a_2\hat{\bf x}$
introduces a compression of the electron guiding
centers along the direction of the channels (i.e. along $\hat{\bf y}'$), 
hence the finite energy cost of such a deformation.
By contrast, a rigid rotation with displacement vectors of the form
${\bf u}_{nm} = \theta\big[ na_1\hat{\bf y} - (n+2m)a_2\hat{\bf x}\big]$ 
leads to displaced lattice vectors (to order $\theta^2$) of the form:
${\bf r}_{nm} = na_1\hat{\bf x}' + (n+2m)a_2 \hat{\bf y}'$,
which corresponds to a perfectly ordered anisotropic WC in the $(x'y')$ frame.
The above analysis clearly shows that, while for a conventional liquid crystal (or stripe)
with a uniform density along the stripes a deformation of the form ${\bf u}({\bf r})=-\theta y{\bf x}$
corresponds to a rigid rotation of the system, in our case where translational invariance 
is also broken along the direction of the channels the above displacement field
does {\em not} correspond to a rigid rotation, and in fact introduces a compression of the 
electron guiding centers along the stripes with a finite elastic energy cost.

The elastic matrix of the AWC has elements given by
$\Phi_{xx}({\bf q})=c_{11}(q)q_x^2 + c_{66}q_y^2 + Kq_y^4$,
$\Phi_{yy}({\bf q})=c_{22}(q)q_y^2 + c_{66}q_x^2$, and 
$\Phi_{xy}({\bf q})=\Phi_{yx}({\bf q})=\big(c_{12}(q)+c_{66}\big)q_x q_y$,
where we have introduced a term $\sim Kq_y^4$ in $\Phi_{xx}$ to account for the bending rigidity of the stripes
at small length scales. In conventional smectics, where the shear modulus $c_{66}$ is exactly zero,
the term $\sim c_{66}(\partial_yu_x)^2$ in the elastic free energy is {\em replaced} by the bending rigidity term
$\sim K(\partial_y^2u_x)^2$ \cite{Chaikin-Lubensky}. In our case, where $c_{66}$ is very small but nonzero,
the elastic energy of the AWC will contain both terms, and phenomena taking place on various length scales
will be governed by one or the other of these two elastic contributions. In particular, one can define
a characteristic wavevector $q_{cr}=\sqrt{c_{66}/K}$ (or, equivalently, a crossover length scale $L_{cr}\sim 1/q_{cr}$)
which delimits the long wavelengths regime ($0 \leq q\ll q_c$) where 
$c_{66}q^2\gg K q^4$ and where the elastic properties of the system are governed by the ordinary elastic 
energy of a two-dimensional lattice, and the short wavelengths regime $q_c < q \leq q_{BZ}$ where the 
elastic properties of the system are smectic-like. Using the value $K=189\,mK$ obtained in Ref. \cite{Wexler2001}, 
and taking $c_{66}\sim 5\times 10^{-5}(e^2/\varepsilon\ell)/A_c$, we obtain the following estimate for the crossover 
length scale:
\begin{equation}
L_{cr} \sim 10\,a\,.
\label{Eq:Lcr}
\end{equation}


As an illustration of the qualitative argument above, in the following we shall discuss the behavior
of the normal modes taking into account the shear modulus $c_{66}$ of the AWC.
In the presence of a magnetic field, the equation of motion for the electron at
lattice site ${\bf R}_i$ can be written in the form (${\bf u}_i$ is the displacement vector of the electron
with respect to its equilibrium position ${\bf R}_i$,
$m^*$ is the effective mass of the electron in
the host semiconductor and $\varepsilon_{\alpha\beta}$ is the totally antisymmetric tensor in two
dimensions)
\begin{equation}
m^*\frac{d^2u_{i\alpha}}{dt^2} = -\sum_{j}\Phi_{\alpha\beta}({\bf R}_i-{\bf R}_j)u_{j\beta}
-\frac{eB}{c}\varepsilon_{\alpha\beta}\frac{du_{i\beta}}{dt}.
\label{eqmot}
\end{equation}
We shall seek a solution to the above equation of motion that
represents a wave with angular frequency $\omega$ and wavevector
${\bf q}$, {\em i.e.} $u_{\alpha}({\bf R}_i)=A_{\alpha}({\bf q})e^{i({\bf q}\cdot{\bf R}_i-\omega t)}$, 
where the ${\bf A}_{\alpha}$'s
are complex coefficients whose ratios specify the relative
amplitude and phase of the vibrations of the electrons within each
primitive cell. Substituting the above expression into
Eq.~(\ref{eqmot}) results in the following secular equation for the
Wigner crystal:
\begin{equation}
\begin{pmatrix}
D_{xx}({\bf q})-\omega^2 & D_{xy}({\bf q})-i\omega\omega_c \\
D_{xy}({\bf q})+i\omega\omega_c  & D_{yy}({\bf q})-\omega^2
\end{pmatrix}
\begin{pmatrix}
A_{x} \\ A_{y}
\end{pmatrix} =0,
\label{secEqWC}
\end{equation}
where $D_{\alpha\beta}$ is the dynamical matrix, which is defined in terms of the elastic matrix
$\Phi_{\alpha\beta}$ by
$D_{\alpha\beta}({\bf q})=\frac{1}{m^*}\sum_{i}{\Phi}_{\alpha\beta}({\bf R}_i)
e^{-i{\bf q}\cdot{\bf R}_i}$.

\begin{figure}[t]
\includegraphics[width=8.09cm, height=5cm]{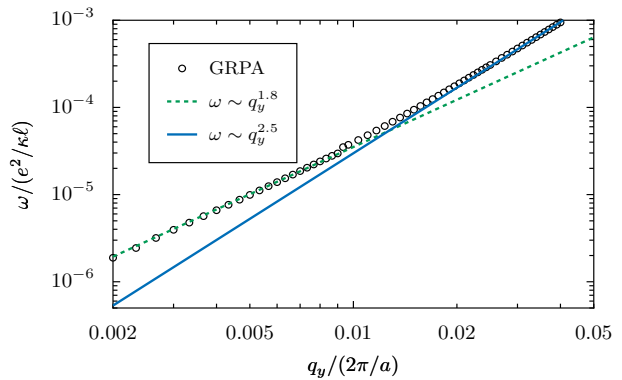}
\caption[]{(Color online) 
Dispersion curve for the stripe state at filling factor $\nu=4.45$. There is a change of
behavior from $\omega\sim q^{1.8}$ to $\omega\sim q^{2.5}$ around $q_{cr}=1/10a$ (in 
the $y$ axis label, $\omega$ is expressed in unit such that $\hbar=1$).
}\label{Fig:dispersion}
\end{figure}

The above secular equation has a non-vanishing solution (for the
${\bf A}_{\alpha}$'s) only if the determinant of the secular matrix
is zero. This leads to the following eigenmodes: \cite{Cote1991}
\begin{equation}
\omega_{\pm}^2({\bf q}) = \frac{1}{2}\Big[\omega_c^2 +
D_{xx}+D_{yy} \pm \sqrt{\Delta({\bf q})}
\Big],
\end{equation}
where
$\Delta({\bf q}) = \omega_c^4 + 2\omega_c^2(D_{xx}+D_{yy}) + (D_{xx}-D_{yy})^2 + 4D^2_{xy}$.
The solutions with a ($-$) sign represent magnetophonon modes, with eigenfrequencies which vanish 
as $q\to 0$, while the solutions with a ($+$) sign correspond to magnetoplasmon modes which tend to $\omega_c$
as $q\to 0$. Solving explicitly for the magnetophonon mode, we obtain in the long wavelength limit where
$c_{11}(q)\simeq c_{22}(q)\simeq c_{12}(q) \equiv c(q)$ the result 
$\omega_{-}^2({\bf q}) =\frac{c(q)c_{66}}{m^2\omega_c^2}\big[q_x^4 - 2 q_x^2q_y^2 + q_y^4]$.
Given that $c_{11}(q)\sim 1/q$, this implies that magnetophonon eigenmodes have the characteristic 
Wigner crystal behavior 
$\omega(q\hat{\bf x})=\omega(q\hat{\bf y})\sim q^{3/2}$ for $q\le q_{cr}$.
On shorter length scales, the bending rigidity term dominates over the shear term.
Neglecting $c_{66}$ in the expression of the elastic matrix, we obtain
$\omega_{-}^2({\bf q})\simeq (Kc_{11}(q)/m^2\omega_c^2)q_y^6$,
with the consequence that $\omega_{-}(q\hat{\bf y})\sim q^{5/2}$ for $q \geq q_{cr}$.
In order to test these predictions, we have performed a direct numerical calculation 
of the dispersion curve using the generalized random phase approximation (GRPA) 
method of Ref. \onlinecite{Cote2003}. The results of this computation are shown 
in Fig. \ref{Fig:dispersion} and seem to indeed confirm the results of the above 
elementary calculation (although the best fit we obtain at small $q$ for a law
of the form $\omega\sim q^\alpha$ is with an exponent $\alpha\simeq 1.8$ instead of $1.5$).


In conclusion, in this paper we have investigated the possibility that the stripe state 
near half filling of higher Landau levels (with index $n\ge 2$) may be anisotropic crystalline states.
By numerically solving the HF equations for a family of oblique realizations of the electronic lattice 
parameterized by an anisotropy parameter $\varepsilon$, with $0\leq \varepsilon < 1$,
and finding the values of $\varepsilon$ that minimize the cohesive
energy of the system, we find within a self-consistent HF approach that the ground state of the system 
in the stripe phase is indeed an anisotropic Wigner crystal, with a small but finite shear modulus.
As an immediate consequence of this, we have shown that the magnetophonon modes have
the usual dispersion of an ordinary Wigner crystal $\omega({\bf q})\sim q^{3/2}$ at wavelengths longer than a crossover 
length scale $L_{cr}$, which we estimate to be of order $10$ lattice spacings near half filling in $n=2$,
with $\omega({\bf q})\sim q^{5/2}$ on shorter length scales.
Previous scale-sensitive predictions for the stripe state, such as predictions of the electromagnetic
response of the stripe state in presence of pinning disorder, will only hold
at length scales shorter than $L_{cr}$, with the standard Wigner crystal behavior
taking over at longer length scales.

\begin{acknowledgments}

The authors acknowledge discussions with R.M. Lewis, L.W. Engel, K. Yang,
M.M. Fogler, and H.A. Fertig.
This work was supported by the National High Magnetic Field Laboratory In 
House Research Program (AME, FDK and ATD), 
by the NSERC of Canada, FQRNT, the Swiss NSF and the NCCR Nanoscience (CBD), and by a grant from the
Natural Sciences and Engineering Council of Canada (RC).

\end{acknowledgments}

\end{document}